\documentstyle[aps]{revtex}
\begin{document}
\twocolumn[ 
{\bf Comment on ``Relativistic Effects of Light in Moving Media with
Extremely Low Group Velocity''}\\
\smallskip
]
In \cite{L&P,L&P:A} Leonhardt and Piwnicki have presented an
interesting analysis of how to use a flowing dielectric fluid to
generate a so--called ``optical black hole''.  Unfortunately there is
subtle misinterpretation in the analysis regarding these ``optical
black holes''. While it is clear that ``optical black holes'' can
certainly exist as theoretical constructs, and while the experimental
prospects for actually building them in the laboratory are excellent,
the particular model geometries of~\cite{L&P,L&P:A} are in fact {\em
not\/} black holes at all.

Leonhardt and Piwnicki consider a vortex geometry, where the
dielectric fluid is swirling in a purely azimuthal direction around a
straight linear core --- there is no radial motion into the core in
their models, and this is enough to prevent the formation of trapped
surfaces and event horizons. This observation depends only on the fact
that the effective metric can globally be cast in the ADM-like form
\begin{equation}
[g_{\mathrm{eff}}]_{\mu\nu} = 
\left( 
\matrix{
{- [c_{\mathrm{eff}}^2 - g_{ab} \; v_{\mathrm{eff}}^a \; v_{\mathrm{eff}}^b ]}
&
{[v_{\mathrm{eff}}]_i}
\cr
{[v_{\mathrm{eff}}]_j}
&
{[g_{\mathrm{eff}}]_{ij}}
}
\right).
\end{equation}
In the acoustic geometry of~\cite{Unruh,Visser}, the constant-time
3-space metric $[g_{\mathrm{eff}}]_{ij}$ is particularly simple (it's
the identity matrix), the effective velocity $[v_{\mathrm{eff}}]_i$
equals the fluid velocity, and $c_{\mathrm{eff}}$ is just the local
speed of sound. In the non-relativistic limit of the non-dispersive
moving-medium optical geometry of~\cite{L&P:A}, the effective velocity
$[v_{\mathrm{eff}}]_i$ equals the fluid velocity adjusted by the
Fresnel drag correction, while $c_{\mathrm{eff}}\to c/n$ is just the
local speed of light, and the 3-metric acquires $O(v^2)$
corrections. In the extreme-dispersion model of~\cite{L&P} the optical
geometry is more complicated but the effective velocity is still
proportional to the fluid velocity. These technical complications do
not affect the key issue: An ergo-region certainly forms once the norm
of this effective velocity exceeds the local effective propagation
speed ($||v_{\mathrm{eff}}|| > c_{\mathrm{eff}}$), but provided that
$c_{\mathrm{eff}}$ remains positive, a trapped surface will form only
if the {\em inward normal component} of the effective velocity exceeds
the local effective propagation speed~\cite{Visser}.  Since there is
no inward velocity in either of the vortex models~\cite{L&P,L&P:A},
there is no possibility of forming an event horizon. This result is
generic and does not depend on any optical-acoustic analogy.

What Leonhardt and Piwnicki actually do is to demonstrate that the
vortex geometries they write down possesses an {\em unstable circular
photon orbit}, very similar in its qualitative properties to the
unstable circular photon orbit that occurs at $r=3M$ in the
Schwarzschild geometry (when the Schwarzschild geometry is written in
down in Schwarzschild coordinates.) Photons/light-rays/null geodesics
with high angular momentum, higher than some critical value which
depends on the details of the fluid velocity profile, either (1) come
in from spatial infinity and return to spatial infinity without ever
crossing the unstable photon orbit, or (2) emerge from the vortex
core, never get past the unstable photon orbit and subsequently fall
back to the vortex core.  Photons/light-rays/null geodesics with lower
than critical angular momentum come in from spatial infinity, cross
the unstable photon orbit, and eventually crash into the vortex
core. If their angular momentum is just fractionally less than
critical they may appear to ``hover'' near the unstable photon orbit.
However, when Leonhardt and Piwnicki use the phrase ``Schwarzschild
radius'' it is the radius of this unstable circular photon orbit they
are referring to, and their usage of the phrases ``Schwarzschild
radius'' and ``event horizon'' has nothing to do with the sense in
which they are defined in general relativity.

Despite this technical issue, which causes problems for the two
particular toy models they discussed~\cite{L&P,L&P:A}, it is clear
that the basic idea is fine --- it certainly is possible to form
``optical black holes'' but only by adding an inward radial component
to the vortex flow. Alternatively, you could think of a nozzle that
accelerates a low-compressibility dielectric fluid to (effectively)
superluminal velocities. Any region of superluminal effective velocity
$v_{\mathrm{eff}}$ will be an ergo-region, and any surface for which
the inward normal component of this effective velocity is superluminal
will be a trapped surface~\cite{Visser}.

If successful, this technique will be able to probe
aspects of {\em semiclassical quantum gravity}, such as the existence
of Hawking radiation.  Because the effective metric is not constrained
by the Einstein equations you only probe {\em kinematic} aspects of
how quantum fields react to being placed on a curved background
geometry, but do not probe any {\em dynamical\/} questions of just how
quantum matter feeds into the Einstein equations to generate real
spacetime curvature~\cite{Visser}.  Though ``effective metric''
techniques are limited in this sense, they are still a tremendous
advance over the current state of affairs.

\bigskip
\noindent
Matt Visser\\
Washington University in Saint Louis\\
Saint Louis, Missouri 63130-4899\\
e-mail: visser@kiwi.wustl.edu\\
homepage: http://www.physics.wustl.edu/\~{}visser


Received 2 Feb 2000; 31 May 2000


\end{document}